\documentclass[11pt, onecolumn]{IEEEtran}
\usepackage{amssymb}
\usepackage{amsmath}
\usepackage{epsfig}

\newtheorem{theorem}{Theorem}

\newcommand{\xr}{\mathbf{r}}

\begin{document}

\date{}
\title{Distributed Data Storage with Minimum Storage Regenerating Codes - Exact and Functional Repair are Asymptotically Equally Efficient }
\author{\authorblockN{Viveck R. Cadambe, Syed A. Jafar, Hamed Maleki\footnote{The ordering of authors is alphabetical.}}\\
\authorblockA{Electrical Engineering and Computer Science\\
University of California Irvine, \\
Irvine, California, 92697, USA\\
Email: {\{vcadambe, syed, hmaleki\}@uci.edu}\\ }}

\maketitle
\begin{abstract}
We consider a set up where a file of size $M$ is stored in $n$ distributed storage nodes, using an $(n,k)$  minimum storage regenerating (MSR) code, i.e., a maximum distance separable (MDS) code that also allows efficient exact-repair of any failed node. The MDS property ensures that the original file can be reconstructed even if  any $n-k$ storage nodes fail. When a node fails, a new node collects data from the remaining $n-1$ healthy nodes and repairs the failed node. The problem of interest in this paper is to minimize the \emph{repair bandwidth} $B$ for \emph{exact} regeneration of the failed node, i.e., the minimum data to be downloaded by the new node to replace the failed node by its exact replica. Previous work has shown that with random network coding, a bandwidth of $B=\frac{M(n-1)}{k(n-k)}$ is necessary and sufficient for functional (not exact) regeneration, i.e., if the repaired new node need not be exactly identical to the failed node, but only information equivalent to it. It has also been shown using interference alignment based techniques that if $k \leq \max(n/2, 3)$ then, surprisingly, there is no extra cost of exact regeneration over functional regeneration and the same repair bandwidth of $\frac{M(n-1)}{k(n-k)}$ suffices for exact regeneration. The practically relevant setting of low-redundancy, i.e., $k/n>1/2$ remains open for $k>3$ and it has been shown that there \emph{is} an extra bandwidth cost for exact repair over functional repair in this case. In this work, we adopt into the distributed storage context an asymptotically optimal interference alignment scheme previously proposed by Cadambe and Jafar for large wireless interference networks.  With this scheme we solve the problem of repair bandwidth minimization for $(n,k)$ exact-MSR codes for all $(n,k)$ values including the previously open case of $k > \max(n/2,3)$. Our main result is that, for any $(n,k)$, and sufficiently large file sizes, there is no extra cost of exact regeneration over functional regeneration in terms of the repair bandwidth per bit of regenerated data. More precisely, we show that $\lim_{M\to \infty} \frac{B}{M} = \frac{n-1}{k(n-k)}$. The result is analogous to the wireless interference channel setting where exact interference alignment through linear beamforming is seen to be infeasible for more than $3$ users, but almost perfect alignment is achieved asymptotically by the Cadambe-Jafar scheme over a large number of signaling dimensions for any number of users.

\end{abstract}

\section{Introduction}
The problem of interest in this paper is to minimize the bandwidth required to exactly repair failed nodes in distributed storage systems. It is well known that maximum distance separable (MDS) codes can be used to reliably store data in distributed storage nodes. To see this, consider a scenario where a file of size $M$ is to be stored in $n$ distributed storage nodes. The file is split into $k$ equal parts of size $M/k$ and stored in the first $k$ storage nodes, also known as systematic nodes. The remaining $(n-k)$ nodes, known as parity nodes or non-systematic nodes, store data of the same size, i.e., $M/k$, adding redundancy to protect from failure of storage nodes. The parity nodes are designed so that a failure of up to $(n-k)$ storage nodes can be tolerated, i.e., the original file can be completely recovered from the data stored at any $k$ nodes out of the original $n$ nodes. Clearly, for this problem, storing the data using an $(n,k)$ MDS code suffices to achieve the required reconstruction criterion, since an MDS code protects the data from $(n-k)$ erasures. Now, consider the case where only $1$ node fails, and a new node is introduced to replace the failed node. The total amount of data to be downloaded by the new node to regenerate a single failed node will be henceforth referred to as the \emph{repair bandwidth}. Clearly, a repair bandwidth of $M$ suffices to repair a failed node since the new node can download data of total size $M$ from any $k$ of the remaining $n-1$ \emph{healthy} nodes to reconstruct the failed node. However, note the inherent inefficiency in the solution - to reconstruct a node of size $M/k$, the newcomer downloads data of size $M$, i.e., $k$ times the size of the data to be repaired. A question of interest is whether this inefficiency is fundamental, or whether the node can be repaired with the new comer downloading data of size less than $M$. More specifically, the question of interest of this paper is \emph{what is the minimum repair bandwidth required to repair a failed node?} The question of minimum repair bandwidth has been studied previously from two perspectives \cite{Dimakis_Godfrey_Wainwright_Ramachandran, Wu_Explicit, Wu_Dimakis, Suh_Ramachandran, Shah_etal}. The first is called functional regeneration \cite{Dimakis_Godfrey_Wainwright_Ramachandran, Wu_Explicit} and the second is called exact (or systematic) regeneration \cite{Wu_Dimakis, Suh_Ramachandran, Shah_etal}.


The functional regeneration problem requires the new node to replace the failed node by a function of the data, so that the reconstructed new node, along with the other nodes satisfy the property of being an $(n,k)$ MDS code. In other words, the repaired node is information equivalent to the originally stored data. Note that in the functional regeneration problem, the data stored by the repaired node need not be identical to the data stored by the failed node; all that is required is that the repaired node along with the other nodes forms a MDS code. This problem has been shown to be equivalent to finding the capacity of a particular wired single-source multi-cast network. Since network coding achieves the cut-set bound in a single-source multi-cast network, the functional regeneration problem has been solved, and it is shown in \cite{Dimakis_Godfrey_Wainwright_Ramachandran} that the minimum bandwidth required is $B = \frac{M(n-1)}{k(n-k)}$. Note that since $\frac{M(n-1)}{k(n-k)}$ is smaller than $M$ (for $k<n$), the solution trivially implies that reconstruction of a single failed node requires a smaller repair bandwidth than reconstruction of $(n-k)$ failed nodes, by a factor of $\frac{(n-1)}{k(n-k)}$.

The focus of this paper is on the exact (or systematic) regeneration problem, where the new comer is required to replace the failed node by a replica, i.e., an identical copy of the failed node. Since exact regeneration ensures that a failed systematic node is replaced by a systematic node, the systematic structure of data storage is retained. There is a practical advantage of preservation of the systematic structure which ensures easy access to the data for a client, since the client can simply download it from the $k$ systematic nodes without any decoding. Note that the constraints for systematic or exact regeneration are stricter than the functional regeneration problem. Since any solution for the exact regeneration problem is also a solution to functional regeneration problem, $\frac{M(n-1)}{k(n-k)}$ serves as a lower bound to the minimum repair bandwidth for the exact repair problem. However, if the repair bandwidth of $\frac{M(n-1)}{k(n-k)}$ suffices, in general, has been an open question. It is this open question that is the focus of this paper. 
\begin{figure*}
\begin{center}
\includegraphics{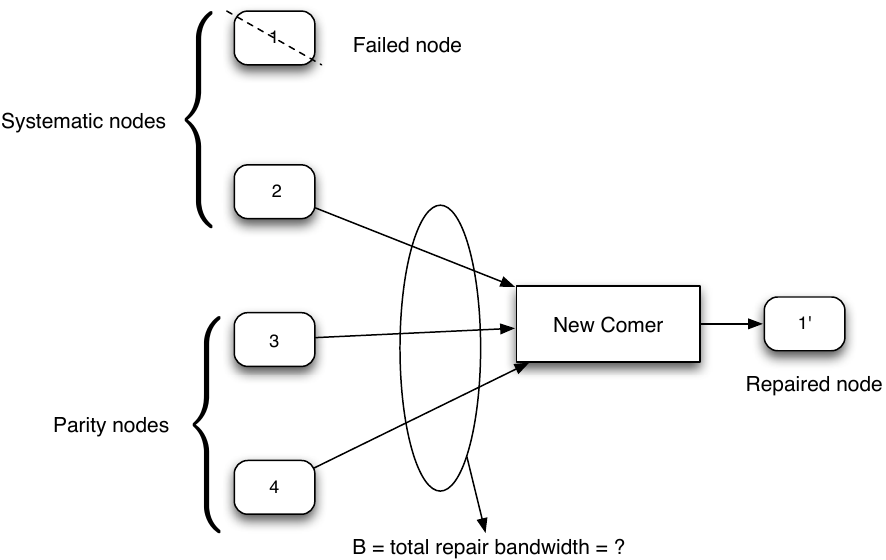}
\caption{Pictorial Representation of Problem Definition for $n=4,k=2$}
\end{center}
\end{figure*}

\subsection{Related Work and Summary of Contributions}
The exact regeneration problem was formulated and solved for the special case of $n=4,k=2$  in \cite{Wu_Dimakis}. The solution was further extended to the more general case of $k\leq \max(3, n/2)$ in \cite{Suh_Ramachandran, Shah_etal}. The results in all these cases yield the same surprising conclusion: there is no price for exact regeneration over functional regeneration, and a repair bandwidth of $\frac{M(n-1)}{k(n-k)}$ suffices even for exact regeneration.  The solution for these cases stems from drawing parallels between the exact regeneration problem and the \emph{wireless} interference channel \cite{Suh_Ramachandran}. Such parallels enable the use of the interference management technique of interference alignment \cite{Jafar_Shamai,MMK} for the exact regeneration problem. However, prior to this work, as far as we are aware, little was known about the minimum repair bandwidth for $k > \max(3,\frac{n}{2})$. From a practical perspective, note that the previously unsolved case of $k > \max(3,\frac{n}{2})$ is important because this case corresponds to the amount of parity data (i.e., number of parity nodes) being smaller than the original file size (number of systematic nodes). This case is briefly studied in reference \cite{Shah_etal} where for $ k > \frac{n}{2}+{1}$, it is shown that the lower bound of $\frac{M(n-1)}{k(n-k)}$ \emph{cannot} be achieved using linear codes. The main contribution of this paper is to make progress in this open problem drawing inspiration from the interference alignment solution for the $K$ user wireless interference channel in \cite{Cadambe_Jafar_int}. We argue that, while that lower bound on the repair bandwidth of $\frac{M(n-1)}{k(n-k)}$ may not be sufficient in general as noted in \cite{Shah_etal}, the repair bandwidth \emph{per bit of repaired data} can indeed achieve this lower bound in the limit of large file sizes for any $k < n$. More precisely, we show that $$\lim_{M\to \infty} \frac{B}{M} = \frac{n-1}{k(n-k)},$$ i.e., for any $k < n$, with sufficiently large amount of data, \emph{there is no cost of exact repair over functional repair in terms of repair bandwidth per bit of repaired data}. An interesting insight of our solution is that the size of the symbol extension in wireless interference channels is analogous to the file size of our solution. Reference \cite{Cadambe_Jafar_int} shows that the optimal number of degrees of freedom of the interference channel cannot be achieved with finite symbol extensions (using linear schemes), but can only be achieved asymptotically in the limiting case of arbitrarily large symbol extensions. This is analogous to our result combined with that of reference \cite{Shah_etal}. The reference shows that the bound of $\frac{M(n-1)}{k(n-k)}$ cannot be achieved exactly. Here, we complete the analogy between file size and symbol extensions in our main result by showing that while the bound is not exactly achievable, it is achievable asymptotically in the limit of large file sizes. We state our main result formally below.
\begin{theorem}
\label{thm:main}
Consider any tuple $(n,k)$ such that $n > k$. For a file of size $M$ stored in $n$ distributed storage nodes as a part of a $(n,k)$ MDS code, the minimum repair bandwidth $B$ for exact regeneration of a (single) failed node satisfies 
$$\lim_{M \to \infty} \frac{B}{M} = \frac{n-1}{k(n-k)}.$$
Equivalently, we can write 
$$ B = \frac{M(n-1)}{k(n-k)} + o(M)$$
\end{theorem}
Our approach to the problem is different from those of \cite{Shah_etal, Suh_Ramachandran} which use Cauchy matrices for code construction for $k < \max(3,n/2)$. Before we proceed to prove the theorem, we take a closer look at a simple case of $(n=4,k=2)$, initially presented in \cite{Wu_Dimakis}, to describe the role of interference alignment in exact regeneration.

\section{The Role of Interference Alignment in Exact Regeneration : $n=4, k=2$}

\begin{figure*}
\begin{center}
\includegraphics{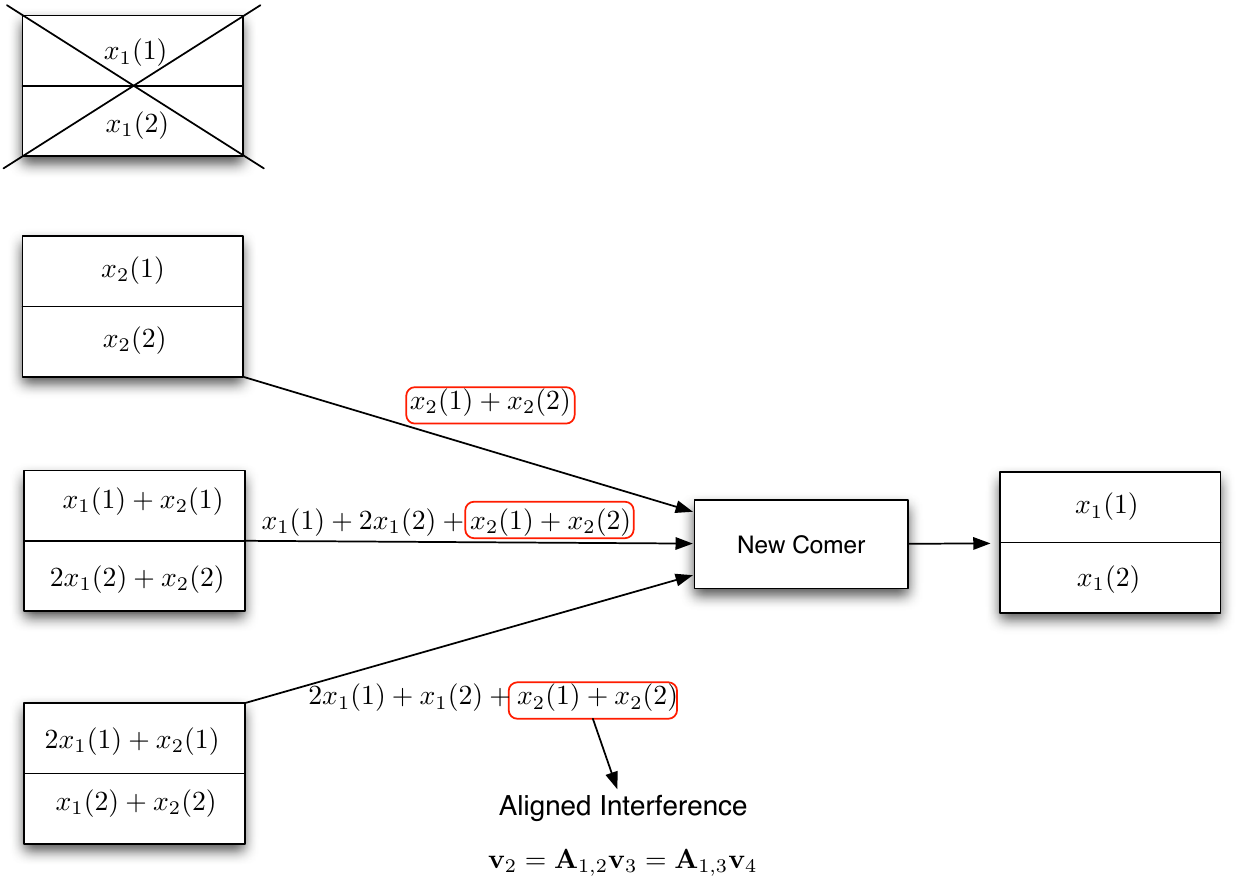}
\caption{ Alignment based Exact Repair for $n=4,k=2$ \cite{Wu_Dimakis}}
\label{fig:Wu_Dimakis}
\end{center}
\end{figure*}
Consider the case where the $n=4,k=2$. Further, let the file to be stored be $\left[\mathbf{x}_1~~\mathbf{x}_2\right]^T$ where $\mathbf{x}_i,i=1,2$ are $2\times1$ vectors over a finite field of size $q$ denoted by $\mathbb{F}_q$.

\subsubsection*{Remark 1} 
Note that for sufficiently large file sizes, the field size $q$ is a design parameter. For the solution for $n=4,k=2$ presented here, any prime $q \geq 3$ suffices. Also by defining $\mathbf{x}_1, \mathbf{x}_2$ to be $2 \times 1$ vectors, we are assuming that the file-size $M=4$, i.e., 4 scalars over the field. For large file sizes, $M$ can be treated as a design parameter, since a code for a specified $M$ can be used for larger files by splitting the file into portions of size $M$.

As in Figure \ref{fig:Wu_Dimakis}, the first systematic node stores $\mathbf{x}_1$ and the second systematic node stores $\mathbf{x}_2$, Let the parity nodes $m=3,4$ store vectors of the form $\mathbf{A}_{m,1}\mathbf{x}_1+\mathbf{A}_{m,2} \mathbf{x}_{2}$, where $\mathbf{A}_{m,i}, i=1,2, m=3,4$ are $2 \times 2$ matrices. 
Now, consider the case where 
$ \mathbf{A}_{3,2} = \mathbf{A}_{4,2}=I$  and $$\mathbf{A}_{3,1}=\left[\begin{array}{cc} 1 & 0 \\ 0 & 2\end{array}\right]~~~\mathbf{A}_{4,1}=\left[\begin{array}{cc} 2 & 0 \\ 0 & 1 \end{array}\right].$$
Note that $\mathbf{x}_{1}, \mathbf{x}_2$ can be reconstructed from any $k=2$ of the $n=4$ nodes. Now, consider the case where the first node fails. First, we present a naive solution which does not align interference. Note that the contents of the node, i.e. $\mathbf{x}_{1}$ can be reconstructed by a new comer downloading $4$ linear combinations (or equations) from any other two nodes. For example, it can be reconstructed by downloading the $2 \times 1$ vector $\mathbf{x}_{2}$ from node $2$ and the $2 \times 1$ vector $\mathbf{A}_{1,1} \mathbf{x}_{1}+\mathbf{A}_{1,2} \mathbf{x}_{2}$ from node $3$. In this case, note that among the $4$ dimensions (corresponding to the $4$ linear equations) at the new comer, two dimensions are occupied by the data to be reconstructed, i.e., the desired data $\mathbf{x}_1$ and two dimensions are occupied by the undesired data or \emph{interference} $\mathbf{x}_2$. 

However, a more efficient solution exists. By \emph{aligning} the interference into $1$ dimension, we can see that the node can be repaired by downloading only $3$ linear combinations of the stored data from the remaining \emph{healthy} nodes. To see this, consider the case where the new comer downloads a total of $3$ linear combinations of the stored data, one from each remaining healthy node, as follows.
\begin{itemize}
\item $\mathbf{v}_{2}^T \mathbf{x}_{2}$ from node $2$ 
\item $\mathbf{v}_{3}^{T} \left( \mathbf{A}_{3,1} \mathbf{x}_1+ \mathbf{A}_{3,2}\mathbf{x}_{2}\right) $ from node $2$ 
\item $\mathbf{v}_{4}^T \left( \mathbf{A}_{4,1} \mathbf{x}_1+ \mathbf{A}_{4,2}\mathbf{x}_{2}\right)$ from node $3$ 
\end{itemize} 
where $\mathbf{v}_{2}=\mathbf{v}_{3}=\mathbf{v}_{4} = [1~~1]^T$ (See Figure \ref{fig:Wu_Dimakis}). As shown in Figure \ref{fig:Wu_Dimakis}, since 
$$\mathbf{v}_2 = \mathbf{A}_{1,2}\mathbf{v}_3 = \mathbf{A}_{1,3}\mathbf{v}_{4}$$
the interference aligns into $1$ dimension at the new comer. Further, since 
$$ \mbox{rank}([\mathbf{A}_{3,1}^T \mathbf{v}_{3}~~\mathbf{A}_{4,1}^T \mathbf{v}_{4}]) = \mbox{dimension of }\mathbf{x}_1 = 2,$$
and $\left[\mathbf{A}_{3,1}^T \mathbf{v}_{3}~~\mathbf{A}_{4,1}^T \mathbf{v}_{4}\right]$ is linearly independent of the aligned interference $\mathbf{v}_2$, the data storage node can reconstruct the desired two dimensional vector $\mathbf{x}_1$  from the received $3$ dimensional vector. The code can be shown to exactly repair any failed node with a repair bandwidth of $3$, i.e., with the new comer collecting $3$ linear combinations from the healthy nodes. We now proceed to extend this for general values of $(n,k)$ and prove our main result.

\section{Proof of Theorem \ref{thm:main}}
We begin by generalizing the setting described in the previous section. The total data is represented by the $M/k \times k$ dimensional matrix $\left[ \mathbf{x}_1~~\mathbf{x}_2~~\ldots~~\mathbf{x}_k\right]$, where $\mathbf{x}_i$ is an $M/k \times 1$ dimensional vector stored by systematic node $i \in \{1,2,\ldots,k\}$. Node $j$, where $j\in \{k+1,k+2,\ldots, n\}$ being a parity node stores the $M/k \times 1$ vector $\mathbf{A}_{j,1} \mathbf{x}_1+\mathbf{A}_{j,2} \mathbf{x}_2+\ldots+\mathbf{A}_{j,k} \mathbf{x}_k$, where $\mathbf{A}_{j,i}$ is a $M/k \times M/k$ square matrix for $i \in \{1,2,\ldots,k\}.$ 
Henceforth, we assume that for $j \leq k$, $$\mathbf{A}_{j,i} = \left\{ \begin{array}{cc}\mathbf{0} & j \neq i\\ \mathbf{I} & j = i\end{array}\right., \forall i \in \{1,2,\ldots,k\}. $$ The above assumption implies that the data stored in node $j \in \{1,2,\ldots, n\}$ is the $M/k \times 1$ vector 
$$ \mathbf{D}_{j} = \sum_{i=1}^{k} \mathbf{A}_{j,i} \mathbf{x}_{i}.$$
Note that $\mathbf{A}_{j,i}$ for $j=k+1,k+2, \ldots, n$ are a design choice that define the code; these matrices will henceforth be referred to as the \emph{coding matrices}. We need to choose these matrices so that the code is an MDS code, i.e., using any subset of $k$ nodes, the entire $M \times 1$ vector of data must be reconstructable. Thus, we need to ensure that 
\begin{equation}\mbox{rank}\left(\left[ \begin{array}{cccc} \mathbf{A}_{j_1, 1}& \mathbf{A}_{j_1, 2}& \ldots &\mathbf{A}_{j_1, k}\\
\mathbf{A}_{j_2, 1}& \mathbf{A}_{j_2, 2}& \ldots & \mathbf{A}_{j_2, k}\\
\vdots& \vdots & \ddots & \vdots\\
\mathbf{A}_{j_k, 1}& \mathbf{A}_{j_k, 2}& \ldots & \mathbf{A}_{j_k, k} \end{array} \right] \right) = M \label{eq:MDS} \end{equation}
for any distinct $j_1, j_2, \ldots, j_k \in \{1,2,\ldots, n\}$.

Now, when a node fails, the new comer collects a $\beta \times 1$ vector from each of the remaining $(n-1)$ healthy nodes where $\beta=\frac{B}{n-1}$, so that the total repair bandwidth is $B$. Our goal is to find the coding matrices $\mathbf{A}_{j,i}, (j, i) \in \{k+1, k+2, \ldots, n\} \times \{ 1,2,\ldots, k\}$ and design the $\beta \times 1$ vector to be downloaded by the new comer so as to meet the required bound (presented in the statement of the theorem). We now describe our solution assuming that a systematic node fails. We will later describe how the solution can be adapted to repair failures of parity nodes. Without loss of generality, let us assume that node $1$ fails.
We provide a linear solution to this problem, so that the $\beta \times 1$ vector downloaded by the new comer from node $j>1$ to repair node $1$ is $\mathbf{V}_{j}^{T} \mathbf{D}_{j} $, where $\mathbf{V}_{j}$ is a $M/k \times \beta$ matrix. The matrices $\mathbf{V}_{j}$ will be henceforth referred to as the repair vectors.
 The new comer now has to regenerate the $M/k \times 1$ vector $\mathbf{x}_1$ using $(n-1)$ vectors of the form $\mathbf{V}_{j}^T \mathbf{D}_j, j=2,3,\ldots,n-1$, each of dimension $\beta \times 1$.  Notice that the $(k-1)$ vectors (of dimension $\beta \times 1$) downloaded using the $(k-1)$ systematic nodes do not contain any information about the desired vector $\mathbf{x}_1$ and can be interpreted as interference. Therefore, the new comer has, apart from the interference, $(n-k)$ vectors of dimension $\beta \times 1$ containing linear combinations of the desired data. Thus, the vectors available at the new comer can be described as follows.
\begin{itemize}
\item $(k-1)$ vectors of the form $\mathbf{V}_j^{T} \mathbf{x}_j,1 < j \leq k$ - these vectors are downloaded from the $(k-1)$ healthy systematic nodes. They contain no information about the desired data, and will be used to cancel interference.
\item $(n-k)$ vectors of the $\mathbf{V}_j^{T} \sum_{i=1}^{k} \mathbf{A}_{j,i} \mathbf{x}_i, k < j \leq n$ - these vectors contain both the desired signal and components of the interference.
\end{itemize}

The goal of our solution will be to completely cancel the interference from the latter $(n-k)$ vectors using the former $(k-1)$ vectors listed above, and then to regenerate $\mathbf{x}_1$ using the latter $(n-k)$ vectors. In order to completely cancel the interference related to $\mathbf{x}_i$ using $\mathbf{V}_i^T \mathbf{x}_i$ by linear techniques, we will need, $\forall j=k+1, k+2, \ldots, n$, and for some $\beta \times \beta$ matrix $\Lambda_j$,
\begin{eqnarray} 
\mathbf{V}_j^T \mathbf{A}_{j,i} \mathbf{x}_i &=& \Lambda_j \mathbf{V}_{i}^{T} \mathbf{x}_i \label{eq:alignment0} \\
\Rightarrow \mbox{rowspan}(\mathbf{V}_j^T \mathbf{A}_{j,i}) &\subseteq& \mbox{rowspan}(\mathbf{V}_i^T), \label{eq:alignment1}\\
\Rightarrow \mbox{colspan}(\mathbf{A}_{j,i}^T \mathbf{V}_{j}) &\subseteq& \mbox{colspan}(\mathbf{V}_i), i=2,3,\ldots,k  \label{eq:alignment}\end{eqnarray}
where (\ref{eq:alignment1}) follows from the fact that the matrices $\mathbf{A}_{j,i}$ and $\mathbf{V}_{i}$ are picked independent of the data $\mathbf{x}_i$, and therefore need to satisfy (\ref{eq:alignment0}) for any data vector $\mathbf{x}_i$. 

 While the above condition ensures that the entire interference can be cancelled, we also need to ensure that, on interference cancellation, the $(n-k)$ vectors of dimension $\beta \times 1$ are sufficient to reconstruct $\mathbf{x}_1$. Note that after interference cancellation, each of the $(n-k)$ vectors is of the form $\mathbf{V}_j^{T} \mathbf{A}_{j,1} \mathbf{x}_1$, for $j=k+1, k+2, \ldots, n$. For linear reconstruction, we need
\begin{eqnarray*}
\mathbf{\Psi} \left[\begin{array}{c} \mathbf{V}_{k+1}^T \mathbf{A}_{k+1,1} \mathbf{x}_1 \\ \mathbf{V}_{k+2}^T \mathbf{A}_{k+2,1} \mathbf{x}_1 \\ \vdots \\ \mathbf{V}_{n}^T \mathbf{A}_{n,1} \mathbf{x}_1 \end{array} \right] &=& \mathbf{x}_1\\
\Rightarrow  \mathbf{\Psi} \left[\begin{array}{c} \mathbf{V}_{k+1}^T \mathbf{A}_{k+1,1}  \\ \mathbf{V}_{k+2}^T \mathbf{A}_{k+2,1} \\ \vdots \\ \mathbf{V}_{n}^T \mathbf{A}_{n,1} \end{array} \right] \mathbf{x}_1 &=& \mathbf{x}_1\end{eqnarray*}
for some $M/k \times (n-k)\beta$ matrix $\mathbf{\Psi}$. Therefore, we need
\begin{equation} \mbox{colspan}(\left[\mathbf{A}_{k+1,1}^T \mathbf{V}_{k+1}~~\mathbf{A}_{k+2,1}^T \mathbf{V}_{k+2}~~\ldots~~\mathbf{A}_{n,1}^T \mathbf{V}_{n}\right]) = \frac{M}{k} \label{eq:reconstruction}\end{equation}

Therefore, our goal is to design $\mathbf{A}_{j,i}$ and $\mathbf{V}_{l}$ for $j\in \{k+1, k+2,\ldots, n\}, i \in \{1,2,\ldots,k \}, l=2,3,\ldots,n$ so that 
\begin{itemize}
\item The code is a $(n,k)$ MDS code.
\item The interference is aligned appropriately so that it can be completely cancelled.
\item The desired signal $\mathbf{x}_1$ can be regenerated at the new comer.
\end{itemize}
Thus, essentially we need to pick $\mathbf{A}_{j,i}$ and $\mathbf{V}_{l}$ for $j\in \{k+1, k+2,\ldots, n\}, i \in \{1,2,\ldots,k \}, l\in\{2,3,\ldots,n\}$ so that (\ref{eq:MDS}), (\ref{eq:alignment}) and (\ref{eq:reconstruction}) are satisfied. Further, as noted in Remark 1, the field size $q$ and $M$ are also design choices (for large file sizes) that we can use to satisfy these conditions. 

\subsection{The solution : Choosing $\mathbf{A}_{j,i},\mathbf{V}_l, M$ and $q$}
For $k \leq \max(3, n/2)$, the solutions of \cite{Suh_Ramachandran, Shah_etal} design these matrices using Cauchy matrices to satisfy these conditions. Here, note that the conditions (\ref{eq:alignment}), (\ref{eq:reconstruction}) are similar to the interference alignment conditions in the interference channel \cite{Cadambe_Jafar_int}. Specifically, (\ref{eq:alignment}) is analogous to the condition that all the interference must align in the $K$-user interference channel, and (\ref{eq:reconstruction}) is similar to the condition that the desired signal must be linearly independent for linear decoding in the interference channel \cite{Cadambe_Jafar_int}. These parallels with enable us to build a solution based on the asymptotically perfect interference alignment scheme of the same reference. 

On noting that there are $\Gamma=(n-k)(k-1)$ alignment equations in (\ref{eq:alignment}), like in \cite{Cadambe_Jafar_int}, we choose $M=k(n-k)\Delta^{\Gamma}$ and $\beta=(\Delta+1)^\Gamma,$ where $\Delta \geq 1$ can be any integer\footnote{The intuition for these choices of $M$ and $\beta$ will hopefully become clear later in this section for a reader unfamiliar with \cite{Cadambe_Jafar_int}.}. For any value of $\Delta$, we show the existence of a field size $q$, matrices $\mathbf{A}_{j,i}, i\in\{1,2,\ldots,k\}, j \in \{k+1,k+2,\ldots,n\}$ and $\mathbf{V}_{l},l\in\{2,3,\ldots,n\}$ so that (\ref{eq:MDS}), (\ref{eq:alignment}), (\ref{eq:reconstruction}) are satisfied and the failed node can be repaired. Finally, we show that our code can be used to repair non-systematic nodes as well. 
Before we proceed to give a random coding based construction of the coding matrices and repair vectors, we will evaluate the repair bandwidth achieved by our scheme.  Noting that our construction is applicable for any value of $\Delta$, we can make $\Delta$, a design parameter, arbitrarily large. As $\Delta\to \infty$, we have $M \to \infty$ and 
$$\lim_{M \to \infty} \frac{B}{M} = \lim_{\Delta\to\infty} \frac{(n-1)(\Delta+1)^{\Gamma}}{k(n-k)\Delta^{\Gamma}} = \frac{(n-1)}{k(n-k)}$$.

We now proceed to explain our construction of coding matrices and repair vectors satisfying the constraints of repair (\ref{eq:MDS}), (\ref{eq:alignment}),  (\ref{eq:reconstruction}).  Our solution, unlike those in references \cite{Suh_Ramachandran, Shah_etal}, is a random coding solution. Specifically, we choose the coding matrices $\mathbf{A}_{j,i}, i \in \{1,2,\ldots,k\}, j \in \{k+1,k+2,\ldots,n\}$ randomly. We then provide an expression for $\mathbf{V}_{l},l\in\{2,3,\ldots,n\}$ as a (random) function of $\mathbf{A}_{j,i}$ so that (\ref{eq:alignment}) is satisfied. Then we show for large field size $q$, that (\ref{eq:MDS}) and (\ref{eq:reconstruction}) are satisfied with a non-zero probability. This implies that there exists at least one choice of coding matrices $\mathbf{A}_{j,i}$ so that all the desired condtions, i.e., (\ref{eq:MDS}), (\ref{eq:alignment}), (\ref{eq:reconstruction}) are satisfied.


\subsubsection*{Design of Coding Matrices, $\mathbf{A}_{j,i}$}
The alignment constraints, (\ref{eq:alignment}), are similar to the alignment constraints for the interference channel (See equation (50) in \cite{Cadambe_Jafar_int}). Note that the matrices $\mathbf{A}_{j,i}$ play a role analogous to channel matrices in wireless interference channels \cite{Cadambe_Jafar_int}. Drawing inspiration from \cite{Cadambe_Jafar_int}, we choose the $M/k \times M/k$ dimensional matrices $\mathbf{A}_{j,i} \forall j=k+1, k+2, \ldots, n$ to be random \emph{diagonal} matrices with each diagonal entry of each matrix chosen independently and uniformly distributed over the non-zero elements of the field $F_q$. In other words, we choose 
\begin{equation}
\mathbf{A}_{i,j}=\left[ \begin{array}{cccc}
a_{i,j}^1& 0& \ldots & 0\\
0& a_{i,j}^2& \ldots & 0\\
\vdots& \vdots & \ddots & \vdots\\
0& 0& \ldots & a_{i,j}^\frac{M}{k} \end{array} \right]
 \end{equation}
with all the diagonal entries chosen independent of each other and independent of all the diagonal entries of all other coding matrices, i.e., with $a_{i,j}^{m}$ chosen independent of $a_{\tilde{i},\tilde{j}}^{\tilde{m}}$ from the non-zero elements of the field, for all $i \neq \tilde{i}$ or $j \neq \tilde{j}$ or $m \neq \tilde{m}$, where $i, \tilde{i} \in \{1,2,\ldots,k\}$, $j, \tilde{j} \in \{k+1,k+2,\ldots,n\}$ and $m,\tilde{m} \in \{1,2,\ldots,\frac{M}{k}\}$. Note that all the coding matrices are full rank since all the diagonal elements are non-zero. We later show that that this code is an MDS code with non-zero probability.  
\subsubsection*{Design of Repair Vectors, $\mathbf{V}_{l}$}
Here, we provide a set of repair vectors that satisfy (\ref{eq:alignment}). We first set the columns of vectors $\mathbf{V}_{l}$ (which are analogous to beamforming vectors in interference channels) 
$$\mathbf{V}_2=\mathbf{V}_3=\ldots=\mathbf{V}_k = \mathbf{V}^{'}$$
$$\mathbf{V}_{k+1}=\mathbf{V}_{k+2}=\ldots=\mathbf{V}_n = \mathbf{V}$$
where, $\mathbf{V}$ and $\mathbf{V}^{'}$ are $M/k \times \beta$ dimensional matrices. 
Then the relations (\ref{eq:alignment}) can be re-written as 
\begin{equation}  \mbox{colspan}(\mathbf{A}_{j,i} \mathbf{V}) \subseteq \mbox{colspan}(\mathbf{V}^{'}), i=2,3,\ldots,k \label{eq:alignment3}\end{equation}
for $j=k+1, k+2,\ldots,n.$ Note that there are $(k-1)(n-k)=\Gamma$ conditions contained in (\ref{eq:alignment3}). We wish to find $\mathbf{V}, \mathbf{V}^{'}$ so that all these conditions are satisfied.

{\it Intuitive understanding of asymptotic alignment: }Before we provide precise expressions for $\mathbf{V}, \mathbf{V}^{'}$, we will intuitively explain the extent of alignment required to to satisfy (\ref{eq:alignment}), (\ref{eq:reconstruction}). Since our bandwidth is restricted by $\beta$, we need $\mbox{rank}(\mathbf{V}) \leq \beta=(\Delta+1)^{\Gamma}$ and $\mbox{rank}(\mathbf{V}^{'}) \leq (\Delta+1)^{\Gamma}$. Further, noting that (\ref{eq:reconstruction}) implies $\sum_{j=k+1}^{n} \mbox{rank}(\mathbf{V}_{j}) \geq \frac{M}{k}$, we get $\mbox{rank}(\mathbf{V}) \geq \frac{M}{k(n-k)}=\Delta^{\Gamma}$. Therefore $\mathbf{V}$ must have at least $\Delta^{\Gamma}$ non-zero linearly independent columns. In order to satisfy (\ref{eq:alignment3}), the span of the $\Gamma \Delta^{\Gamma}$ non-zero column vectors of the matrix
$$\left[ \mathbf{A}_{k+1,1} \mathbf{V}~~~~\mathbf{A}_{k+2,1} \mathbf{V}~~~~\ldots~~~~\mathbf{A}_{n,1} \mathbf{V}~~~~\mathbf{A}_{k+1,2} \mathbf{V}~~~~\mathbf{A}_{k+2,2} \mathbf{V}~~~~\ldots~~~~\mathbf{A}_{n,k} \mathbf{V}\right]$$
should align in the space spanned by the $(\Delta+1)^{\Gamma}$ column vectors of $\mathbf{V}^{'}$. For large values of $\Delta$, since $\frac{\Delta^{\Gamma}}{ (\Delta+1)^{\Gamma}}\rightarrow 1$, and all the coding matrices have a full rank of $M/k$, we have $\frac{\mbox{rank}(\mathbf{A}_{j,i} \mathbf{V})}{\mbox{rank}(\mathbf{V}^{'})}\rightarrow 1$ for any $j \in \{k+1,\ldots,n\}, i \in \{1,2,\ldots,k\}$. From (\ref{eq:alignment3}) this implies that $\mbox{colspan}(\mathbf{A}_{j,i}\mathbf{V})\approx \mbox{colspan}(\mathbf{V}^{'})$. In other words, the alignment between the $\Gamma$ matrices on the left hand side of the $\Gamma$ relations indicated by (\ref{eq:alignment3}) is asymptotically perfect for large $\Delta$. Next we return to the mathematical construction of the alignment scheme.

Following the arguments of \cite{Cadambe_Jafar_int, Cadambe_Jafar_reflection}, we choose the set of non-zero column vectors of $\mathbf{V},\mathbf{V}^{'}$ as shown below\footnote{For convenience, we ignore the abuse in notation of these equations; the quantity on the left denotes the matrix, whereas the quantity on the right only denotes the set of non-zero columns of the matrix.}, 
\begin{eqnarray}\mathbf{V}=\left\{\left(\prod_{ \substack{
j=k+1,\ldots,n \\
i=2,\ldots,k}} \mathbf{A}_{j,i}^{\alpha_{j,i}}\right)\mathbf{w}: \alpha_{k+1,2},...,\alpha_{n,k}\in\left\{0,1,...,\Delta-1\right\}\right\} \label{eq:V2}\\
\mathbf{V}'=\left\{\left(\prod_{ \substack{
j=k+1,...,n \\
i=2,...,k}} \mathbf{A}_{j,i}^{\alpha_{j,i}}\right)\mathbf{w}: \alpha_{k+1,2},...,\alpha_{n,k}\in \left\{0,1,2,...,\Delta\right\}\right\} \label{eq:V1}
\end{eqnarray}
where the entries of the $M/k \times 1$ column vector $\mathbf{w}$ are chosen uniformly over the non-zero elements of the field and independent of all the coding matrices.

Thus, the elements of $\mathbf{V}$ contain products of (diagonal) coding matrices  corresponding to interference symbols contained in the parity nodes, with each matrix raised to an exponent that is allowed to take integer values from $0$ upto $\Delta-1$. Since there are $\Gamma=(k-1)(n-k)$ coding matrices and $\Delta$ possible distinct values for the exponent of each matrix, the total number of elements, i.e. column vectors,  in $\mathbf{V}$ is $\Delta^\Gamma$. Similarly, the total number of column vectors in $\mathbf{V}^{'}$ is $(\Delta+1)^\Gamma$. To understand the notation better, consider, e.g.,  the case where $\Delta=1$. Then, $\mathbf{V}=\mathbf{w}$, i.e., just one column vector,  and $\mathbf{V}^{'}$ contains all the $2^{\Gamma}$ vectors of the form 
$$\mathbf{A}_{k+1,2}^{\alpha_{k+1,2}} \mathbf{A}_{k+1,3}^{\alpha_{k+1,3}}\ldots \mathbf{A}_{k+1,n}^{\alpha_{k+1,n}} \mathbf{A}_{k+2,2}^{\alpha_{k+2,2}} \ldots \mathbf{A}_{n,k}^{\alpha_{n,k}} \mathbf{w}$$
where $\alpha_{j,i} \in \{0,1\}$. For any general value of $\Delta$, the columns of $\mathbf{V}$ are of the form
$$\mathbf{A}_{k+1,2}^{\alpha_{k+1,2}} \mathbf{A}_{k+1,3}^{\alpha_{k+1,3}}\ldots \mathbf{A}_{k+1,n}^{\alpha_{k+1,n}} \mathbf{A}_{k+2,2}^{\alpha_{k+2,2}} \ldots \mathbf{A}_{n,k}^{\alpha_{n,k}} \mathbf{w}$$
where $\alpha_{j,i} \in \{0,1,\ldots, \Delta-1\}$ and $\mathbf{V}^{'}$ has columns of the form
$$\mathbf{A}_{k+1,2}^{\alpha_{k+1,2}} \mathbf{A}_{k+1,3}^{\alpha_{k+1,3}}\ldots \mathbf{A}_{k+1,n}^{\alpha_{k+1,n}} \mathbf{A}_{k+2,2}^{\alpha_{k+2,2}} \ldots \mathbf{A}_{n,k}^{\alpha_{n,k}} \mathbf{w}$$
where $\alpha_{j,i} \in \{0,1,\ldots,\Delta\}$. Note that the ordering of the matrices $\mathbf{A}_{j,i}$ in the above notation is irrelevant, since the coding matrices, being diagonal, commute. This commuting property is the key to the alignment scheme. Because the ordering of matrices is irrelevant, it is readily verified that multiplying any column vector from $\mathbf{V}$ by any of the $\mathbf{A}_{j,i}$ involved, produces a column vector contained in $\mathbf{V}^{'}$. This is because multiplication by $\mathbf{A}_{j,i}$ simply raises the corresponding exponent of the element in $\mathbf{V}$ by one, but the elements of $\mathbf{V}^{'}$ already include all such terms. Since the set of columns of $\mathbf{A}_{j,i} \mathbf{V}$ is a sub-set of the columns of $\mathbf{V}^{'}$ for any $j\in\{k+1,k+2,\ldots,n\}, i \in \{2,3,\ldots,k\}$, it is evident that  this choice of repair vectors satisfies (\ref{eq:alignment3}), and equivalently, (\ref{eq:alignment}).
\subsubsection*{Proof of (\ref{eq:MDS}), (\ref{eq:reconstruction})}
We have now chosen coding matrices and repair vectors so that the alignment constraints (\ref{eq:alignment}) are satisfied. We now need to show (\ref{eq:MDS}) and (\ref{eq:reconstruction}). In order to show that the matrices of (\ref{eq:MDS}) and (\ref{eq:reconstruction}) are full rank, it is enough to show that their determinants are non-zero. Notice that the determinant of the matrix of (\ref{eq:MDS}), i.e., 
\begin{equation}
\left[ \begin{array}{cccc} 
\mathbf{A}_{j_1, 1}& \mathbf{A}_{j_1, 2}& \ldots &\mathbf{A}_{j_1, k}\\
\mathbf{A}_{j_2, 1}& \mathbf{A}_{j_2, 2}& \ldots & \mathbf{A}_{j_2, k}\\
\vdots& \vdots & \ddots & \vdots\\
\mathbf{A}_{j_k, 1}& \mathbf{A}_{j_k, 2}& \ldots & \mathbf{A}_{j_k, k} \end{array} \right] 
\end{equation}
is a polynomial in its entries. Note that there are $\binom{n}{k}$ polynomials of this kind, which can be represented, for $l=1,2,\ldots,\binom{n}{k}$, as $$f_{l}:\mathcal{A} \rightarrow \mathbb{F}_{q},$$ where $$\mathcal{A} = \left\{a_{j,i}^{m}: j\in\{k+1,k+2,\ldots,n\}, i \in \{1,2,\ldots,k\}, m \in \{1,2,\ldots, (n-k)\Delta^{\Gamma}\}\right\} \rightarrow \mathbb{F}_{q},$$ denotes all the diagonal entries of the coding matrices. In the appendix, we show that each of these polynomials is a non-zero polynomial. 

Similarly, we need to show (\ref{eq:reconstruction}), i.e., 
$$
\mbox{colspan}(\left[\mathbf{A}_{k+1,1} \mathbf{V}~~\mathbf{A}_{k+2,1} \mathbf{V}~~\ldots~~\mathbf{A}_{n,1} \mathbf{V}\right]) = \frac{M}{k} $$
where $\mathbf{V}$ has $\Delta^{\Gamma}=\frac{M}{k(n-k)}$ non-zero columns chosen using (\ref{eq:V2}). Using these non-zero columns (i.e., discounting the columns of $\mathbf{V}$ which are zero) the above matrix is of dimension $\frac{M}{k}\times \frac{M}{k}$. Therefore, to show that this square matrix has a full rank of $\frac{M}{k}$, we need to show that its determinant is non-zero. Since a determinant is a polynomial function of its entries, the determinant expansion above is a polynomial $$g:\mathcal{A} \cup \{w_1,w_2,\ldots,w_{(n-k)\Delta^{\Gamma}}\}\rightarrow \mathbb{F}_{q},$$ where $\mathbf{w}=[ w_1~~w_2~~\ldots~~w_{(n-k)\Delta^{\Gamma}}]^T$. An argument very similar to Lemma 1 of \cite{Cadambe_Jafar_X} can be used to show that the polynomial formed by this matrix for our solution is a non-zero polynomial (See also Appendix III in \cite{Cadambe_Jafar_int}). Thus, the product $f_1(.)f_2(.)...f_{\binom{n}{k}}(.) g(.)$ is non-zero polynomial of $\mathcal{A} \cup \{w_{1}, w_{2},\ldots,w_{(n-k)\Delta^{\Gamma}}\}$. Using Schwartz-Zippel Lemma, for large enough $q$, we have at least one choice of coding matrices and repair vectors such that these polynomials do not evaluate to non-zero, and therefore a solution exists so that (\ref{eq:MDS}),(\ref{eq:reconstruction}) are satisfied.
\subsubsection*{Repair of Non-Systematic (Parity) nodes}
So far, we have discussed an achievable scheme for regenerating a systematic node. The code $\mathbf{A}_{j,i}$ constructed here can also be used to regenerate a failed parity node in the same manner. To see this, suppose that a parity node, say node $k+1$, fails. The new comer intends to regenerate $\mathbf{D}_{k+1}$. Let 
$$\mathbf{x}_1^{'}=\mathbf{D}_{k+1}=\sum_{i=1}^{k} \mathbf{A}_{k+1,i} \mathbf{x}_{i}.$$ Since the code is an MDS code, using a change of basis, we can write 
\begin{eqnarray*} \mathbf{x}_1 &=& \mathbf{A}^{'}_{1,1} \mathbf{x}_1^{'} +\sum_{i=2}^{k} \mathbf{A}^{'}_{1,i} \mathbf{x}_i \\
\mathbf{x}_j &=& \mathbf{A}^{'}_{j,1} \mathbf{x}_1^{'} +\sum_{i=2}^{k} \mathbf{A}^{'}_{j,i} \mathbf{x}_i, j=k+2,k+3,\ldots,n\\
\end{eqnarray*}
where $\mathbf{A}^{'}_{j,i}$ are all diagonal. In other words, a change of basis can essentially transform the regeneration of a parity node to appear like regeneration of a systematic node, i.e., with nodes $2,3,\ldots,k+1$ viewed as systematic nodes storing data $\mathbf{x}_2,\mathbf{x}_3,\ldots,\mathbf{x}_k,\mathbf{x}_1^{'}$; nodes $1,k+2,k+3,\ldots,n$ are viewed as parity nodes using coding matrices $\mathbf{A}_{j,i}^{'}, j \notin \{2,3,\ldots,k+1\}$. Since all the coding matrices are diagonal, the problem can be solved in a similar manner as above, i.e, the vectors downloaded by the new comer can be constructed as in (\ref{eq:V1}),(\ref{eq:V2}) and be verified to satisfy a property similar to (\ref{eq:alignment}). The only thing that remains is to verify if a reconstruction criterion similar to (\ref{eq:reconstruction}) is satisfied. In order to show this, it is enough to show that all the (random) diagonal entries of any new coding matrix $\mathbf{A}_{j,i}^{'}$ for $j \notin \{2,3,\ldots,k+1\}$ are uniformly distributed over the non-zero entries of the field which are independent of each other and independent of all diagonal entries in the other new coding matrices, much like our original construction. Showing this independence property will ensure that our earlier proof of (\ref{eq:reconstruction}) is applicable. In order to show this independence property, we explicitly evaluate $\mathbf{A}_{j,i}^{'}$ for $j \notin \{2,3,\ldots,k+1\}$ as follows.
\begin{eqnarray}
\mathbf{A}^{'}_{j,i} = \left\{\begin{array}{cc}
\mathbf{A}_{k+1,1}^{-1} & j=1, i=1\\
-\mathbf{A}_{k+1,1}^{-1} \mathbf{A}_{k+1,i} & j=1, i\in\{2,3,\ldots,k\}\\
\mathbf{A}_{j,i} \mathbf{A}_{k+1,1}^{-1}& j\in\{k+2,\ldots,n\}, i =1\\
\mathbf{A}_{j,i} -\mathbf{A}_{k+1,1}^{-1} \mathbf{A}_{k+1,i} & j\in\{k+2,\ldots,n\}, i \in \{2,3,\ldots,k\}\end{array}\right.\label{eq:non-systematic}
\end{eqnarray}
Now, note that if $\mathbf{S}_1$ and $\mathbf{S}_2$ are two diagonal matrices with their diagonal entries drawn independently and uniformly distributed over the non-zero elements of the field, then each of the matrices $\mathbf{S}_1\mathbf{S}_2, \mathbf{S}_1+\mathbf{S}_2, -\mathbf{S}_2$ and $\mathbf{S}_2^{-1}$ has diagonal entries uniformly distributed over the non-zero elements of the field. Further each of these matrices are independent of $\mathbf{S}_1$. This implies that all the diagonal entries of $\mathbf{A}_{j,i}^{'}, j \notin \{2,3,\ldots,k+1\}$ are distributed independently of each other, and uniformly distributed over the non-zero elements of $\mathbf{F}_q$. Also, this property can be used in (\ref{eq:non-systematic}) to verify that all the entries of any coding matrix $\mathbf{A}_{j,i}^{'}$ are independent of the entries of all the entries of any other new coding matrix $\mathbf{A}_{\tilde{j},\tilde{i}}^{'}$ for $i \neq \tilde{i}$ or $j \neq \tilde{j}$.  For example, $\mathbf{A}_{1,1}^{'}$ is independently distributed of $\mathbf{A}_{1,2}^{'}$ since the entries of $\mathbf{A}_{k+1,2}$ are independent of $\mathbf{A}_{k+1,1}$ in our original code construction. Thus, the basis transformation preserves the required independence criteria and a property similar to (\ref{eq:reconstruction}) holds. This completes the proof.

\section{Conclusion}
We have shown that, per bit of data to be reconstructed, surprisingly, there is no loss of exact regeneration over functional regeneration in terms of the amount of repair bandwidth per bit of repaired data, in the limit of large file sizes, regardless of the desired redundancy level. The result is in contrast with  previous work in \cite{Shah_etal} where it is shown that there is an efficiency loss for exact regeneration over functional regeneration especially for low redundancy levels. However, note that the two results do not contradict each other. While our asymptotic alignment scheme can approach arbitrarily close to the cut-set bound on minimum repair bandwidth per bit of repaired data, the bound is not achieved with exact equality. Also unlike previous work in \cite{Suh_Ramachandran, Shah_etal} we do not provide explicit codes or specify the minimum field size, since our arguments are based on properties of random matrices. Directions for ongoing work include interference alignment solutions for exact repair for each point on the storage-bandwidth tradeoff curve.

\appendix
\section{Proof of (\ref{eq:MDS})}
\label{app:1}
We intend to show that the determinant of the matrix in (\ref{eq:MDS}) is a non-zero polynomial in its entries. Assuming, without loss of generality, that $j_1, j_2, \ldots, j_k$ are in ascending order, let 
$j_1, j_2, \ldots, j_{k-m} \in \{1,2,\ldots, k\}$ and $j_{k-m+1}, j_{k-m+2}, \ldots, j_{k} \in \{k+1,k+2,\ldots,n\}$. Therefore, we need to show that the determinant of the following matrix is a non-zero polynomial of its entries.
$$
\left[ \begin{array}{cccc} 
\mathbf{A}_{1, 1}& \mathbf{A}_{1, 2}& \ldots &\mathbf{A}_{1, k}\\
\mathbf{A}_{2, 1}& \mathbf{A}_{2, 2}& \ldots & \mathbf{A}_{2, k}\\
\vdots& \vdots & \ddots & \vdots\\
\mathbf{A}_{k-m, 1}& \mathbf{A}_{k-m, 2}& \ldots & \mathbf{A}_{k-m, k}\\
\mathbf{A}_{k+1, 1}& \mathbf{A}_{k+1, 2}& \ldots & \mathbf{A}_{k+1, k}\\
\mathbf{A}_{k+2, 1}& \mathbf{A}_{k+2, 2}& \ldots & \mathbf{A}_{k+2, k}\\
\vdots& \vdots & \ddots & \vdots\\
\mathbf{A}_{k+m, 1}& \mathbf{A}_{k+m, 2}& \ldots & \mathbf{A}_{k+m, k} \end{array} \right] 
$$
Since $$\mathbf{A}_{j,i} = \left[ \begin{array}{cc}\mathbf{0} & j \neq i\\ \mathbf{I} & j = i\end{array}\right], \forall i \in \{1,2,\ldots,k\} $$ 
we want the following matrix to be full rank. 
\begin{equation}
\left[ \begin{array}{ccccc}
\mathbf I_{\frac{M}{k}\times \frac{M}{k}}& \ldots & \mathbf 0_{\frac{M}{k}\times \frac{M}{k}}& \ldots &\mathbf 0_{\frac{M}{k}\times \frac{M}{k}}\\
\mathbf 0_{\frac{M}{k}\times \frac{M}{k}}& \ddots & \mathbf 0_{\frac{M}{k}\times \frac{M}{k}}& \ldots &\mathbf 0_{\frac{M}{k}\times \frac{M}{k}}\\
\mathbf 0_{\frac{M}{k}\times \frac{M}{k}}& \ldots & \mathbf I_{\frac{M}{k}\times \frac{M}{k}}& \ldots &\mathbf 0_{\frac{M}{k}\times \frac{M}{k}}\\
\mathbf{A}_{k+1, 1}& \ldots &\mathbf{A}_{k+1, k-m}& \ldots & \mathbf{A}_{k+1, k}\\
\vdots& \vdots & \vdots & \ddots & \vdots\\
\mathbf{A}_{k+m, 1} & \ldots & \mathbf{A}_{k+m,k- m}& \ldots & \mathbf{A}_{k+m, k} \end{array} \right] 
\end{equation}
Therefore, we essentially need to show that the determinant formed by the above matrix is non-zero. Since the first $(k-m)\frac{M}{k} \times (k-m)\frac{M}{k}$ matrix is the identity matrix, expanding the determinant along the first $(k-m)\frac{M}{k}$ rows, the determinant can be shown to be equal to the determinant of the following matrix.
\begin{small}
\begin{equation}
\mathbf{P}= \left[ \begin{array}{cccc}
\mathbf{A}_{k+1, (k-m)+1}& \mathbf{A}_{k+1,(k-m)+2}& \ldots & \mathbf{A}_{k+1, k}\\
\mathbf{A}_{k+2, (k-m)+1}& \mathbf{A}_{k+2,(k-m)+2}& \ldots & \mathbf{A}_{k+2, k}\\
\vdots& \vdots & \ddots & \vdots\\
\mathbf{A}_{k+m, (k-m)+1}& \mathbf{A}_{k+m, (k-m)+2}& \ldots & \mathbf{A}_{k+m, k} \end{array} \right] \label{eq:det} 
\end{equation}
\end{small}
We need to show that the determinant of the $m \frac{M}{k} \times m \frac{M}{k}$ matrix $\mathbf{P}$ is non-zero. Note that we have
\begin{equation}
\mathbf{A}_{j,i}=\left[ \begin{array}{cccc}
a_{j,i}^1& 0& \ldots & 0\\
0& a_{j,i}^2& \ldots & 0\\
\vdots& \vdots & \ddots & \vdots\\
0& 0& \ldots & a_{j,i}^\frac{M}{k} \end{array} \right]
\label{diag_matrice}
 \end{equation}
where each $a_{j,i}^{l}$ is independent of $a_{j^{'},i^{'}}^{l^{'}}$ for $i\neq i^{'}$ or $j \neq j^{'}$ or $l \neq l^{'}$. Since interchanging the rows or columns of a matrix does not change its determinant except for its sign, we make the row and column exchange operations to simplify $\mathbf{P}$. Let the rows of $\mathbf{P}$ be $\xr_1, \xr_2, \ldots, \xr_{mM/k}$. Now, we only need to show that the determinant of $\mathbf{P}^{'}$ is non-zero where
$$\mathbf{P}^{'} = \left[\begin{array}{c} \xr_{\pi_1(1)}\\ \xr_{\pi_1(2)} \\ \vdots \\ \xr_{\pi_1(mM/k)} \end{array} \right]$$
where $\pi_1:\{1,2,\ldots,mM/k\} \rightarrow \{1,2,\ldots,mM/k\}$ is a permutation. Now, further, let $\mathbf{c}_1, \mathbf{c}_2, \ldots, \mathbf{c}_{mM/k}$ be the columns of $\mathbf{P}^{'}$. We then perform column exchange operations of $\mathbf{P}^{'}$ to get the matrix $\mathbf{P}^{''} = [ \mathbf{c}_{\pi_2(1)}~~\mathbf{c}_{\pi_2(2)}~~\ldots~~\mathbf{c}_{\pi_2(mM/k)}],$ where, $\pi_2:\{1,2,\ldots,mM/k\} \rightarrow \{1,2,\ldots,mM/k\}$ is also a permutation. Now, that the determinant of  $\mathbf{P}$ is non-zero is equivalent to showing that the determinant of $\mathbf{P}^{''}$ is non-zero. Choosing the permutations $\pi_1, \pi_2$ as 
\begin{eqnarray*}
\pi_1(i)=\pi_2(i) &=& 1+ \left\lfloor \frac{i-1}{M/k} \right\rfloor+\left( i - \left\lfloor\frac{i-1}{M/k}\right\rfloor \frac{M}{k}-1\right) m\\
\end{eqnarray*}
it can be verified that the $m\frac{M}{k} \times m\frac{M}{k}$ matrix $\mathbf{P}^{''}$ has a block diagonal structure, with $\frac{M}{k}$ blocks of size $m \times m$. The $i$th block of $\mathbf{P}^{''}$ is 
\begin{equation}
\left[ \begin{array}{cccc}
a_{k+1, (k-m)+1}^i& a_{k+1, (k-m)+2}^i& \ldots & a_{k+1, k}^i\\
a_{k+2, (k-m)+1}^i& a_{k+2, (k-m)+2}^i& \ldots & a_{k+2, k}^i\\
\vdots& \vdots & \ddots & \vdots\\
a_{k+m, (k-m)+1}^i& a_{k+m, (k-m)+2}^i& \ldots & a_{k+m, k}^i \end{array} \right].
 \end{equation}
Since the determinant of a block diagonal matrix is a product of the determinant of each of its blocks, and the determinant of the square matrix formed by the above block is a non-zero polynomial of its entries, the determinant of the matrix in (\ref{eq:MDS}) is a non-zero polynomial of its entries, as required.

\bibliographystyle{ieeetr}
\bibliography{Thesis}

\end{document}